\documentstyle[12pt,a4]{article}
\begin{document}
\begin{center}
\large {\bf  Atomic Quantum Zeno Effect for Ensembles and Single Systems}

\vspace*{.5cm}
Almut Beige\footnote
{e-mail: beige@theorie.physik.uni-goettingen.de}, 
Gerhard C. Hegerfeldt\footnote
{e-mail: hegerf@theorie.physik.uni-goettingen.de}
and Dirk G. Sondermann\footnote
{e-mail: sonderma@theorie.physik.uni-goettingen.de} \\[.5cm]
\normalsize
   Institut f\"ur Theoretische Physik\\   
   Universit\"at  G\"ottingen\\
   Bunsenstr. 9\\
   D-37073 G\"ottingen, Germany
\end{center}

\begin{abstract}
The so-called quantum Zeno effect is essentially a consequence of the
projection postulate for ideal measurements. To test the effect 
Itano et al.\ have performed an experiment on an ensemble of atoms 
where rapidly repeated level
measurements were realized by means of short laser pulses. Using
dynamical considerations we give an 
explanation why the projection postulate can be applied in good
approximation to such measurements. Corrections to  ideal measurements  
are determined explicitly. This is used to discuss in
how far the experiment of Itano et al.\ can be considered as a test of the
quantum Zeno effect. We also analyze a new 
possible experiment on a single atom where stochastic light and dark
periods can be interpreted as manifestation of the quantum Zeno effect. 
We show that  the measurement point of view gives a
quick and intuitive understanding of experiments of the above type,
although a finer analysis has to take the corrections into account.
\end{abstract}                
\vspace*{0.5cm}

\noindent {\bf 1. Introduction}
\vspace*{0.5cm}

For an ideal measurement of an observable $A$ on a system in 
state $|\psi \rangle$ standard quantum mechanics predicts as
possible outcomes the eigenvalues $ a_i $ of $A$. 
Each $a_i$ is
found with probability $\| I\!\!P_i |\psi \rangle \|^2 $,
where $I\!\!P_i$ is the projector onto the eigenspace belonging to
$a_i$. The projection postulate then states that right after a
measurement for which $a_i$ is found
the system is in the state $ I\!\!P_i |\psi \rangle$. 
The  projection postulate as currently used has been
formulated by L\"uders \cite{Lue}. For observables with 
degenerate eigenvalues his formulation differs from that of  
von Neumann \cite{N}.
L\"uders stressed its  provisional character: ``The projection postulate
will be employed only until a better understanding of the actual
measurement process has been found'' \cite{Lue2}.
As pointed out to us by Sudbury \cite{Sud}, in the first edition of his 
famous book  Dirac \cite{Dirac} defines
observations  causing minimal disturbance. These correspond to
L\"uder's prescription. Curiously though, in later editions this passage has
been omitted. 

One consequence of the projection postulate is, under some mild technical
conditions, the so-called quantum Zeno effect \cite{MiSu}.
It predicts a slow-down of the time development due to rapidly
repeated measurements. If the time between two measurements, 
$\Delta t$, goes to zero the system is frozen on a subspace. 

An experiment to test the quantum Zeno effect was performed by Itano
et al.\ \cite{Wine}. They stored several thousand two-level ions
in a trap. Initially all ions were prepared in the ground state. Then
a pulse of a weak field in resonance was applied which pumped an ion
in the ground state $\left|1\right>$
to the excited state $\left|2\right>$.
Such a pulse is called a 
$\pi$ pulse. During this $\pi$ pulse $n$  population 
measurements of the ion levels were performed. Starting in
level 1 and increasing the number $n$
of measurements, i.e. decreasing the  time $\Delta t$
between two measurements, less  atoms were 
found in level 2 at the end of
the $\pi$ pulse. For $n=1$ to $64$ the results were in good agreement  
with the predictions of the quantum Zeno effect.

Opinions have been divided in the literature about the relevance 
of this experiment for the quantum Zeno effect
(cf.\ e.g.\ Refs.\ \cite{Peres,crit4,crit1,B15,B16,crit5,Gag,bloch}). 
Some have called it a dramatic verification of the
effect. Other maintained that it had
nothing to do with the effect since the measurements were realized by
short laser pulses which should be included in the dynamics. 
Therefore the experiment can be understood either by
Bloch equations \cite{B15,B16,Gag,bloch}, which describe the
interaction of an ensemble of atoms with the laser, or by  
incorporating the laser pulses as an external field in the 
Hamiltonian \cite{crit4}. Thus  one can describe the effect of
the laser pulses in a purely dynamical way without any 
measurement interpretation.

Namiki and collaborators have investigated the quantum Zeno effect
and possible experimental verifications for other systems, in
particular spin systems, and have discussed the connection with
measurement theory \cite{Namiki}. The quantum Zeno effect and the
experiment of Ref.~\cite{Wine} have given rise to a large number of
publications \cite{interest}.

The aim of this paper is twofold. First, we want to discuss the
slightly different roles the quantum Zeno effect plays for an ensemble
as opposed to a single system. Second, drawing on recent results of
ours \cite{BeHe1,BeHeSo,BeHe2}
we will describe a state measurement on the two-level atoms by short
laser pulses  in a way
which, although also dynamical, shows why  the laser pulses 
are so well described by the notion of measurement. 
We will explain how and why the measurement point of view
gives such a quick and intuitive understanding of experiments of the above
type, especially of experiments with {\em single} systems. However,
the description by ideal measurements is only approximate. 
Corrections to ideal measurements are necessary and can be determined 
explicitly. 
\vspace*{0.5cm}

\noindent {\bf 2. Role  of the projection postulate for single
systems and ensembles}

\vspace*{0.5cm}

In principle the projection postulate deals with ideal measurements
on individual systems (selective measurements). 
E.g. for  a single system in state $|\psi \rangle$
one can measure $|\psi\rangle$, i.e.\ the observable
$|\psi\rangle\langle\psi|$, at times
$\Delta t,\,2\Delta t,\,\ldots,\,n\Delta t=t$. Then
the probability to find the system at each measurement in $|\psi\rangle$
until $t$ equals
\begin{eqnarray} \label{P(t)1a}
P(t)& = & \| U(\Delta t) |\psi \rangle \langle \psi | .\,.\,.\,
|\psi \rangle \langle \psi | U(\Delta t) 
|\psi \rangle \langle \psi | U(\Delta t) |\psi \rangle \|^2 \\
& = & 1+ \Delta t \, {t \over\hbar} \left( \langle \psi |H| \psi \rangle^2
-\langle \psi |H^2| \psi \rangle \right)+{\cal O}(\Delta t^2) ~.
\label{P(t)1}
\end{eqnarray}
Here the Hamiltonian is taken to be constant in time and suitable
domain assumptions have been made. In the limit
when the time $\Delta t$ between two measurements goes to zero one has 
\begin{eqnarray} \label{P(t)2}
\lim_{\Delta t \to 0} P(t) &=& 1 ~.
\end{eqnarray} 
In this (idealized) limit the system freezes in its initial state.
For a system described by a finite dimensional Hilbert space
the domain assumptions are always fulfilled.
The general case can be treated with less stringent technical
assumptions \cite{MiSu}.

In an ensemble of systems one may initially prepare all systems in the
same state $|\psi \rangle$. Then the density matrix $\rho(t)$ after $n$
measurements (non-selective measurements) is a mixture (`incoherent
superposition') of various subensembles resulting from selective
measurements on individual systems. Now $P(t)$, as determined in 
Eq.~(\ref{P(t)1}), is equal to the relative
size of the subensemble found in $|\psi \rangle$  at each measurement.
Eq.~(\ref{P(t)2}) shows that for decreasing $\Delta t$  
this subensemble increases and thus the
density matrix remains closer to the initial state for longer times,
i.e. there is an overall slow-down of the time-development of the
density matrix and eventually a freezing.

As an  example we consider  ideal state measurements on a two-level 
system \cite{Cook1,Cook2}. Both levels are assumed to be
stable. The transition between level 1 and 2 is driven by a weak field
in resonance. This leads to so-called Rabi oscillations in which 
the atomic state oscillates continuously between $|1\rangle$ and $|2\rangle$. 
In an appropriate interaction picture the time development operator reads
\begin{eqnarray} \label{u(t)}
U(t) &=& \cos {\Omega_2\over2}t
-{\rm i} \sin {\Omega_2\over2}t\,
\left(|1 \rangle \langle 2|+|2 \rangle \langle 1| \right) ~,
\end{eqnarray}
where the so-called Rabi frequency $\Omega_2$ is proportional to the
amplitude of the driving field. Now one can perform rapidly repeated 
ideal measurements of  $|1\rangle\langle1|$ or $|2\rangle\langle2|$ at
times $\Delta t$ apart. Then the probability to find another
measurement result than before is given by 
\begin{equation} \label{p12}
P(|1\rangle \to |2\rangle) = P(|2\rangle \to |1\rangle) 
= \sin^2 {\Omega_2\over2}\Delta t~,
\end{equation}
which is proportional to $ \Delta t^2$ if $\Omega_2 \Delta t \ll 1$. 
For $|\psi\rangle = |1\rangle$ or $|2\rangle$, Eq.~(\ref{P(t)1a})
becomes
\begin{equation}\label{4a}
P(t) = \cos^{2n} {\Omega_2\over2}\Delta t~.
\end{equation}
This leads to a stochastic path for a single atom as shown in
Fig.~1. Each ideal measurement projects the atom onto state $|1\rangle$
or $|2\rangle$, respectively. Then the increase or decrease of the 
population of level 2 goes initially quadratic in time. As can be seen
in Fig.~1, more frequent measurements lead to a slow-down of the 
time-development of the atomic state. 

Let us imagine that the measurement apparatus emits a light signal each
time the atom is found in $|1\rangle$.
Then one would observe stochastically  
alternating light and dark periods. In Fig.~2 the lines mark
times when the atom is found to be in state $|1\rangle$, with the
accompanying light signal. 
When the time between two measurements becomes smaller, the
periods where the atom is found  without interruptions to be in
$|1\rangle$ becomes longer. This corresponds to a light period. 
The same holds for dark periods, with no light signals. 
The atom seems to stay in one state for
some length of time. As discussed further below this suggests a possible 
experimental demonstration of the quantum
Zeno effect for a single atom. 

 From Eq.~(\ref{p12}) the mean length and the standard deviation of the
light and dark periods can be obtained. One finds, with $t_n =
n\Delta t$,
\begin{eqnarray}\label{meana}
\overline T_{\rm L}  =  \sum^\infty_{n=1}~n \Delta t [P(t_{n-1}) 
-P(t_n)] 
\end{eqnarray}
and the same expression for $\overline T_{\rm D}$. Thus, with Eq.
(\ref{4a}), one easily obtains
\begin{equation} \label{mean}
\overline T_{\rm L} ~=~ \overline T_{\rm D} =
\frac{\Delta t}{\sin^2 {1 \over 2} \Omega_2 \Delta t}~.
\end{equation}
The standard deviation is found as
\begin{equation}\label{dev}
\Delta T_{\rm L} ~=~ \Delta T_{\rm D} = 
\Delta t\,\frac{\cos {1\over 2} \Omega_2 \Delta t}
{\sin^2 {1 \over 2} \Omega_2 \Delta t}~.
\end{equation}
For small $\Delta t$ one has a very broad distribution (see Fig.~3). 
In this case
of ideal measurements there is a symmetry between light and dark periods.

When measuring repeatedly on an ensemble (``gas'') of atoms  and
assuming a light signal for each individual atom found in $|1\rangle$
would have a statistical overlap of individual light and dark
periods, and the individual periods would be no longer discernible.
One would just have a decrease of the overall luminosity, from which
one could deduce the above mentioned slow-down in the time
development of the density matrix.  
\vspace*{0.5cm}

\noindent {\bf 3. Analysis of a measurement proposal }

\vspace*{0.5cm}

Cook \cite{Cook1} made a proposal how to measure whether the atom is in
state $|1\rangle$ or $|2\rangle$. His idea was to use an auxiliary,
rapidly decaying third level and a short strong laser pulse. 
Occurrence and absence 
of  fluorescence was taken as indication that the atom was found
in state $|1\rangle$ and $|2\rangle$, respectively. That fluorescence
yields the state $|1\rangle$ is quite easy to understand.
Let us first  assume that, as shown in Fig.~4, the 1--2 transition is not 
driven (no $\pi$ pulse) when the strong laser pulse is applied.
Then one can argue as follows.
If the laser pulse produces fluorescence then after 
the last photon emitted during the laser pulse 
the atom is in the ground state. Until the end of the laser pulse 
the atom is again driven into a superposition of level 1 and 3. 
Then the $|3\rangle$ component decays in a short transient time 
$\tau_{\rm tr}$. Hence shortly after the end of the laser pulse
the atom is in state $|1 \rangle$. 
If the atom is initially in state $|2\rangle$ no photons are emitted,
because a transition to level 3 is not possible. Thus for the total
system (atom plus radiation field) the time development until after
the end of the laser pulse should transform 
\begin{eqnarray*}
|1\rangle\,|\mbox{vacuum}\rangle \qquad &\mbox{into}& \qquad
|1\rangle\,|\mbox{photons}\rangle\\
|2\rangle\,|\mbox{vacuum}\rangle \qquad &\mbox{into}& \qquad
|2\rangle\,|\mbox{vacuum}\rangle
\end{eqnarray*}
where the state $|\mbox{photons}\rangle$ contains practically no
vacuum part. By linearity one then has for an arbitrary initial
state  the transformation
$$\bigl(\alpha_1  |1\rangle + \alpha_2 |2\rangle\bigr)\,|\mbox{vacuum}\rangle 
  \quad\longmapsto\quad \alpha_1  |1\rangle |\mbox{photons}\rangle
  + \alpha_2 |2\rangle |\mbox{vacuum}\rangle \,.$$

Complications arise if the 1--2 transition is driven. 
In how far Cooks proposal \cite{Cook1} corresponds to an ideal 
measurement and state reduction can be analyzed in detail by means of the 
quantum jump approach \cite{HeWi,Wi,He,HeSo}, which is essentially 
equivalent to quantum trajectories \cite{QT} and to the Monte-Carlo
wave function approach \cite{MC}. 

With the quantum jump approach one can show that an atom evolves with a
so-called 
conditional time development operator $U_{\rm cond}$ as long as no
photons are detected. Thus an atom, at time $t_0$
in the  initial state 
\begin{equation} \label{psi}
|\psi \rangle = \alpha_1 |1\rangle +\alpha_2 |2\rangle ~,
\end{equation}
is at time $t$ in the state
\begin{equation}
|\psi^0(t) \rangle 
= U_{\rm cond}(t,t_0) |\psi \rangle
\equiv \exp \left( {-{{\rm i}\over \hbar}(t-t_0) H_{\rm cond}}\right)
|\psi\rangle~,
\end{equation}
if no photons are found between $t_0$ and $t$. The conditional Hamiltonian 
$H_{\rm cond}$ can be explicitly calculated. In
the case of a system as  in Fig.~5 one has, in the same
interaction picture as in Eq.~(\ref{u(t)}),
\begin{equation} \label{hcond}
H_{\rm cond}={\hbar \over 2} \left[ \sum_{i=2}^3 \Omega_i \left( 
|1 \rangle \langle i|+ |i \rangle \langle 1|  \right) 
-{\rm i} A_3 |3 \rangle \langle 3| \right]~.
\end{equation}
The conditional Hamiltonian is non-Hermitian and the norm of 
$|\psi^0(t) \rangle $ decreases in time. 
The probability to find no photon between $t_0$ and $t$ is given by
\begin{eqnarray*}
P_0(t,\psi) = \| |\psi^0 (t) \rangle \|^2
= \| U_{\rm cond}(t,t_0) |\psi \rangle \|^2 ~.
\end{eqnarray*}
The probability density for the emission of the first photon in the 
interval $[t,t+{\rm d}t]$ is equal to $w_1(t,\psi) {\rm d}t$ with  
\begin{eqnarray*} 
w_1(t,\psi) = -{{\rm d} \over {\rm d}t} P_0(t,\psi) ~.
\end{eqnarray*}
It is also shown in the quantum jump approach that a system is
in state $|1\rangle$ after the emission of a photon
if its level structure is as in Figs.~4 and 5.

This can now be used to analyze the effect of the laser pulse on an
atom in more detail. If a laser pulse produces no fluorescence, the
state of the atom at the end of the laser pulse is given by
$$|\psi^0 (\tau_{\rm p}) \rangle 
  = \exp \left( -\frac{\rm i}{\hbar} H_{\rm cond} \tau_{\rm p} \right)
  |\psi \rangle ~.$$
If we consider first the case where no field is applied to the
1--2 transition ($\Omega_2=0)$, 
the eigenvalues $\lambda_i $, $i=1,2,3$, of $H_{\rm cond}$ are easily
calculated. One of them, denoted by $\lambda_2$, is zero and the
corresponding eigenvector is $|2\rangle$. The two remaining
eigenvalues have negative imaginary part. If all eigenvalues are
pairwise different (otherwise one has to take limits) one can write  
\begin{equation} \label{exp} 
|\psi^0 (\tau_{\rm p}) \rangle 
= \sum_{i=1}^3 \exp \left( -\frac{\rm i}{\hbar} \lambda_i 
\tau_{\rm p} \right) 
|\lambda_i \rangle \langle \lambda^i|\psi \rangle~,
\end{equation}
where the $ |\lambda_i \rangle $'s are the eigenvectors of the 
conditional Hamiltonian and the $ \langle \lambda^i| $'s the reciprocal
vectors, with $\langle \lambda^j|\lambda_i \rangle = \delta_{ij}$. 
If $\tau_{\rm p}$ is large enough the exponentials with $i=1$ and $i=3$ in
Eq.~(\ref{exp}) have dropped off to zero and thus at the end of the laser
pulse the state of an atom without any emission has become
\begin{equation}
|\psi^0 (\tau_{\rm p}) \rangle = \alpha_2 |2 \rangle~.
\end{equation}
As shown in Refs.~\cite{BeHe1,BeHeSo} this is valid if the duration of
the laser pulse satisfies the condition
\begin{equation} \label{lengthp}
\tau_{\rm p} \gg \mathop{\rm max} \left\{1/A_3,\,A_3/\Omega_3^2
\right\} ~.
\end{equation}
Then the probability to find no photon equals $|\alpha_2|^2$, and 
the atom is in $|2\rangle$ at the end of the laser pulse. Analogously a 
condition on the transient time $\tau_{\rm tr}$ can be determined. 
The transient time has to be long enough to allow the vanishing of 
possible $|3 \rangle$ components,
\begin{equation} \label{lengthtr}
\tau_{\rm tr} \gg 1/A_3 ~.
\end{equation}
 
Summarizing the discussion so far, if $\Omega_2 = 0$ 
the effect of the laser pulse can be interpreted as a
projection of the atom onto  states $|1\rangle$ or $|2\rangle$,
respectively, if conditions (\ref{lengthp}) and (\ref{lengthtr}) are 
satisfied. This happens with the same probabilities as predicted for
an ideal measurement. The resulting state is characterized by occurrence,
or no occurrence, of a burst of light. The time this takes is equal to 
$\tau_{\rm p}+\tau_{\rm tr}$, the sum of laser pulse length and the
transient time.

If the weak field is not turned off when the laser pulse is on, as
shown in Fig.~5, some complications arise. The weak field can cause
a small additional transition between levels 1 and 2 while the strong
laser pulse is on, as well as  
during the transient time. This leads to small
corrections, which we have explicitly determined in Refs.~\cite{BeHe1,BeHeSo}
by means of the quantum jump approach. One has three 
parameters which have to be small to make the following interpretation 
possible, namely 
\begin{equation}
\epsilon_{\rm A}=\frac{\Omega_2}{A_3}\ll 1,~  
\epsilon_{\rm R}=\frac{\Omega_2}{\Omega_3}\ll 1~{\rm and}~  
\epsilon_{\rm p}=\frac{\Omega_2 A_3}{\Omega_3^2}\ll 1~.
\end{equation} 
Now  one can say that if the condition in Eq.~(\ref{lengthp}) is 
satisfied and if the time between two laser pulses 
is longer than the transient
time in Eq.~(\ref{lengthtr}) then the laser pulse ``projects'' 
the atom onto states $\rho^>_{\rm P} \approx |1 \rangle \langle 1|$
and $\rho^0_{\rm P} \approx |2 \rangle \langle 2|$, respectively.  
This happens with nearly the same probabilities as predicted by
projection postulate for an ideal measurement of the 
states $|1\rangle$ or $|2\rangle$.

This will now be discussed in more detail. We assume the atom to be in
an arbitrary initial state which may also have a $|3\rangle$ component,
\begin{eqnarray} \label{psiallg}
|\psi \rangle &=& \alpha_1|1\rangle+\alpha_2|2\rangle+\alpha_3|3\rangle~.
\end{eqnarray}
The state of an atom which does not emit any photon while the laser
pulse is on can be determined by using the conditional Hamiltonian 
and Eq.~(\ref{exp}). The atom now practically evolves into the eigenstate 
of $H_{\rm cond}$ for the eigenvalue with smallest imaginary part.
This will be denoted by $|\lambda_2\rangle$, and it has also
a $|3\rangle$ component. An elementary calculation gives
\begin{equation}\label{i7}
\lambda_2 = ~\frac{\Omega_2}{2}\,\epsilon_{\rm p}\,( 1 + {\cal O}
({\mbox{\boldmath $\epsilon$}}^2 ))
\end{equation}
and 
\begin{eqnarray} \label{i11}
| \lambda_2 \rangle &=& - {\rm i} \epsilon_{\rm p} |1 \rangle + | 2 \rangle -
\epsilon_R | 3 \rangle + {\cal O} ({\mbox{\boldmath $\epsilon$}}^2)~ 
\\ \label{i12}
| \lambda^2 \rangle &=& \hphantom{-} {\rm i} \epsilon_{\rm p} | 1 \rangle
+ | 2 \rangle -
\epsilon_R | 3 \rangle + {\cal O} ({\mbox{\boldmath $\epsilon$}}^2)~.
\end{eqnarray}
At the end of the laser pulse the state of an atom without emissions 
has thus evolved into
\begin{equation}\label{i14}
 {\rm e}^{-\lambda_2 \tau_{\rm p}}\langle \lambda^2 |
\psi \rangle \cdot | \lambda_2 \rangle
\end{equation}
and the probability for this is the norm squared,
\begin{eqnarray} \label{p0}
\!\!\!\!\!\!P_0(\tau_{\rm p};\psi)\!\!\!&=&\!\!\! 
(1-\epsilon_{\rm p}\Omega_2\tau_{\rm p})|\alpha_2|^2
+2 \epsilon_{\rm p} {\rm Im}(\alpha_1\overline{\alpha}_2)
-2 \epsilon_{\rm R} {\rm Re}(\alpha_2\overline{\alpha}_3)
+{\cal O}(\mbox{\boldmath $\epsilon$}^2) \\ \nonumber
&\approx&\!\!\! |\alpha_2|^2 ~.
\end{eqnarray}
Thus at the end of a laser pulse the normalized state of an atom
without fluorescence is given by
\begin{eqnarray} \label{rho0}
\rho^0(\tau_{\rm p}) \equiv |\lambda_2\rangle \langle \lambda_2|  
= \left( \begin{array}{ccc}
0 & -{\rm i}\epsilon_{\rm p} & 0 \\
{\rm i}\epsilon_{\rm p} & 1 & -\epsilon_{\rm R} \\
0 & -\epsilon_{\rm R} & 0 \end{array} \right)
+{\cal O}(\mbox{\boldmath $\epsilon$}^2)~.
\end{eqnarray}

To obtain the density matrix which describes an atom which does emit a
burst of light during the laser pulse one has to average over all possible
ways how photons can be emitted. This has been done for 
$\Omega_3^2 \ll A_3^2$ in Ref.~\cite{BeHe1} and for the general case in
Ref.~\cite{BeHeSo}. Right at the end of a laser pulse 
an atom with emissions is shown to be in the (normalized) state 
\begin{eqnarray} \label{rho>}
\rho^>(\tau_{\rm p}) &=& 
\left( \begin{array}{ccc}
A_3^2+\Omega_3^2 & {\rm i}\epsilon_{\rm p}A_3^2 & {\rm i}A_3\Omega_3 \\
-{\rm i}\epsilon_{\rm p}A_3^2 & \epsilon_{\rm p}\Omega_2 \tau_{\rm  p}
A_3^2 & \epsilon_{\rm R}(A_3^2+\Omega_3^2) \\
- {\rm i}A_3\Omega_3 & \epsilon_{\rm R}(A_3^2+\Omega_3^2) & 
\Omega_3^2 \end{array} \right) \nonumber \\
&& \times (A_3^2+2\Omega_3^2+\epsilon_{\rm p}\Omega_2 \tau_{\rm p}A_3^2)^{-1}
\;+\;{\cal O}(\mbox{\boldmath $\epsilon$}^2)
\end{eqnarray}
which has non-negligible $|3\rangle$ components and which is also independent 
from the initial state $|\psi \rangle$ of the atom.
(Strictly speaking, Eq.\ (\ref{rho>}) holds only if the leading contribution
$1-|\alpha_2|^2$ of $1-P_0(\tau_{\rm p};\psi)$,
the probability to detect photons, does not vanish or becomes itself
${\cal O}(\mbox{\boldmath $\epsilon$})$.
In these exceptional cases $\rho^>$ {\em does} depend on the
initial state $|\psi\rangle$.)

After the end of the laser pulse the $|3\rangle $ components decay
during the transient time. Simultaneously the 1--2 transition is weakly
pumped. For times long enough for the third level contributions 
to have vanished 
it has been shown in Ref.~\cite{BeHeSo} that a time $\tau$
after the end of the laser pulse an atom without 
any emission is in the state 
\begin{equation} \label{nullP}
\rho^0(\tau_{\rm p}+\tau) = U(\tau) ~\rho^0_{\rm P}~ U^\dagger(\tau)~,  
\end{equation}
where $U(\tau)$ is the ``free''
time development operator of Eq.~(\ref{u(t)}), which describes the small 
driving of the 2-level atom by the weak field and where $\rho^0_{\rm P}$ is
in the 1--2 subspace and given by 
\begin{equation} \label{rho0P}
\rho^0_{\rm P} = \left( \begin{array}{cc}
0 & -{\rm i}\epsilon_{\rm p} \\
{\rm i}\epsilon_{\rm p} & 1 \end{array} \right)
+{\cal O}(\mbox{\boldmath $\epsilon$}^2)\;\approx\; |2\rangle \langle 2| ~.
\end{equation}
Analogously an atom with emissions is, a time $\tau$ after the end of the
laser pulse, in the (normalized) state 
\begin{equation} \label{.gt.P}
\rho^>(\tau_{\rm p}+\tau) = U(\tau) ~\rho^>_{\rm P}~ U^\dagger(\tau)  
\end{equation}
with
\begin{eqnarray} \label{rho>P}
\rho^>_{\rm P} &=& 
\left( \begin{array}{cc} A_3^2+2\Omega_3^2 
& {\rm i}\epsilon_{\rm p}A_3^2-\frac{\rm i}{2}\epsilon_{\rm A}\Omega_3^2 \\
-{\rm i}\epsilon_{\rm p}A_3^2+\frac{\rm i}{2}\epsilon_{\rm A}\Omega_3^2 
& \epsilon_{\rm p}\Omega_2 \tau_{\rm  p} A_3^2 
\end{array} \right) \nonumber \\
&&\times (A_3^2+2\Omega_3^2+\epsilon_{\rm p}\Omega_2 \tau_{\rm p}A_3^2)^{-1}
\;+\;{\cal O}(\mbox{\boldmath $\epsilon$}^2)
\;\approx\; |1\rangle \langle 1|~.
\end{eqnarray}

Eqs.\ (\ref{nullP}) and (\ref{.gt.P}) suggest the interpretation that,
under the stated conditions, the laser pulse effectively ``projects''
the atom at time $\tau_{\rm p}$ onto the states $\rho^0_{\rm P}$ or
$\rho^>_{\rm P}$, respectively, which then undergo a ``free'' time
development.
The probabilities to find the atom in
$\rho^0_{\rm P}$ or $\rho^>_{\rm P}$
are nearly the same as for an ideal measurement of the states
$|1\rangle$ or $|2\rangle$.
However, in particular in this case the laser pulse is not quite an ideal
measurement, and corrections have been given in the above formulas.
\vspace*{0.5cm}

\noindent {\bf 4. Ensembles: Discussion of the experiment of Itano et al.}

\vspace*{0.5cm}

Now the experiment of Itano et al.~\cite{Wine} to test the quantum
Zeno effect on an ensemble of atoms can be analyzed in more detail. 
As shown in Fig.~6 and discussed in the Introduction they had an
ensemble of atoms in a trap (a gas with negligible cooperative effects) and
applied a weak field for the duration $T_\pi = \pi/\Omega_2$ (a $\pi$
pulse). During this $\pi$ pulse $n$
strong laser pulses of duration $\tau_{\rm p}$ were applied. Initially 
all atoms were prepared in the ground state. Therefore without the
strong laser pulses all atoms would be in the state $|2\rangle$ at the end of 
the $\pi$ pulse.

Every single atom is influenced by the strong laser pulses. Therefore, 
the effect of every laser pulse on the ensemble can be regarded as 
simultaneous and approximately ideal measurements 
on each single atom (with corrections, as discussed in Section 3). 

Itano et al.~determined experimentally the population of level 2
at the end of the $\pi$ pulse for $n$ laser pulses during this time
\cite{Wine}.
Their results are shown in
the last column of Table 1. If one interprets the effect of a laser 
pulse as an ideal and instantaneous measurement one obtains
the first column. Better results are 
obtained by assuming an ideal state reduction and taking the finite
duration of the laser pulse for its realization into account (second
column). 

The results of the last section have been used elsewhere
\cite{BeHe1,BeHeSo} to analytically calculate the population of level
2, i.e. with the proper corrections up to order 
$\mbox{\boldmath$\epsilon$}^2$  to the projection postulate taken
into account. 
With the parameters of the experiment we obtain the
third column of Table 1. As seen in column 4, 
a numerical solution of the corresponding Bloch equation leads to
comparable  results, although it 
does not give the same direct physical insight.

\begin{table}
\noindent \begin{tabular}{rccccc}\hline
&\multicolumn{2}{c}{Projection Postulate}&Quantum&Bloch&Observed\\
$n$&$\Delta t=T_\pi/n$&$\Delta t=T_\pi/n-\tau_{\rm p}$&Jump&equations&
\cite{Wine}\\\hline
1 & 1.00000 & 0.99978 & 0.99978 & 0.99978 & 0.995 \\
2 & 0.50000 & 0.49957 & 0.49960 & 0.49960 & 0.500 \\
4 & 0.37500 & 0.35985 & 0.36062 & 0.36056 & 0.335 \\
8 & 0.23460 & 0.20857 & 0.20998 & 0.20993 & 0.194 \\
16 & 0.13343 & 0.10029 & 0.10215 & 0.10212 & 0.103 \\
32 & 0.07156 & 0.03642 & 0.03841 & 0.03840 & 0.013 \\
64 & 0.00371 & 0.00613 & 0.00789 & 0.00789 & \llap{$-$}0.006 \\
\hline
\end{tabular}
\caption { Predicted and observed population 
of level 2 at the end of the $\pi$ pulse for $n$ laser pulses of 
length $\tau_{\rm p}$. $\Delta t$ is the time between two measurements.}
\end{table}   
\vspace*{0.5cm}

\noindent {\bf 5. A possible experiment on a single atom}

\vspace*{0.5cm}

It should be possible to perform the following experiment on a single 
atom in a trap. The weak field  which drives the 1--2 transition, is
kept 
continuously on and will not be turned off after time $T_\pi$. In
addition, strong laser pulses of 
length $\tau_{\rm p}$  are applied repeatedly at times $\Delta t$
apart, as discussed in Section 3 and  depicted in Fig.~7.

A measurement point of view gives a quick and intuitive 
understanding what to expect, namely a stochastic sequence of 
fluorescence bursts forming light periods alternating with dark periods.
Their duration should increase
with decreasing distance between the laser pulses. 

On the other hand, a 
dynamical point of view can directly employ the results of Section
3. With those results one can determine the
probability to find no emissions during a laser pulse, if there had been a
burst of fluorescence or no photons, respectively,  during the
preceding laser pulse. Eq. (\ref{p12}) is now replaced by
\begin{eqnarray*}
P({\rm emissions} \to {\rm no}~{\rm emissions}) 
& = & P(\rho^>_{\rm P} \to \rho^0_{\rm P}) \equiv p~, \\
P({\rm no}~{\rm emissions} \to {\rm no}~{\rm emissions}) 
& = & P(\rho^0_{\rm P} \to \rho^0_{\rm P}) \equiv q~,
\end{eqnarray*}
provided that $\Delta t$ is not to short,
\begin{equation}\label{Bed}
\Delta t\gg1/A_3\quad\mbox{and}\quad
  \left(\Omega_2\Delta t\right)^2\gg\mbox{\boldmath$\epsilon$}\,.
\end{equation}
The first of these conditions ensures that the $|3\rangle$ components
of $\rho^0$ and $\rho^>$ have enough time to decay completely so that
it is possible to make use of the states
$\rho^0_{\rm P}$ and $\rho^>_{\rm P}$.
If the second condition is violated then the  state at the beginning
of  the {\em first} pulse in a light period 
is very close to $\rho^0 $, and therefore the state $\rho^>$ after
the first pulse has to be calculated with initial state of the form
$\rho^0+{\cal O}(\mbox{\boldmath$\epsilon$})$.
For such a state, however, 
one has $1-P_0={\cal O}(\mbox{\boldmath $\epsilon$})$
so that Eq.~(\ref{rho>}) fails. Thus, if the second
condition in Eq.~(\ref{Bed}) does not hold the first pulse in a light
period has to be treated differently from the rest.

The detailed calculation is given elsewhere \cite{BeHe3}. Under the
above conditions the result is
\begin{eqnarray} \label{pq}
p &=& \sin^2{\Omega_2\over2}\Delta t+\epsilon_{\rm p} 
\left(2s\frac{A^2_3+\Omega_3^2}{A^2_3 + 2 \Omega_3 ^2}
+\frac{1}{2}\Omega_2\tau_{\rm p} c\frac{3 A_3^2 + 2 \Omega_3^2}
{A^2_3 + 2 \Omega^2_3} - \frac{1}{2} \Omega_2 \tau_{\rm p}
\right)\nonumber\\
&& -\frac{1}{2} \epsilon_{\rm A} s
\frac{\Omega_3^2}{A^2_3+2\Omega_3^2}+{\cal O}(\mbox{\boldmath$\epsilon$}^2)\\
q &=& \cos^2{\Omega_2\over2}\Delta t-\epsilon_{\rm p} \left( 2s + \frac{1}{2}
\Omega_2\tau_{\rm p}(1 + c)\right)+{\cal O}(\mbox{\boldmath$\epsilon$}^2)
\end{eqnarray}
with $s\equiv\sin\Omega_2\Delta t$ and $c\equiv\cos\Omega_2\Delta t$.

The probability for a period of exactly $n$ consecutive laser pulses
{\em with} fluorescence among all such light periods is $(1 -
p)^{n-1}p$. The mean duration $\overline T_{\rm L}$ of light periods is
therefore
\[
\overline 
T_{\rm L} = \sum^\infty_{n=1} (\tau_{\rm p} + \Delta t)n (1-p)^{n-1} p
\]
which gives
\begin{equation}\label{37}
\overline T_{\rm L}= \frac{\tau_{\rm p} + \Delta t}{p}~.
\end{equation}
Similarly one finds for the dark periods
\begin{equation} \label{38}
\overline T_{\rm D} = \frac{\tau_{\rm p} + \Delta t}{1 - q}~.
\end{equation}
These results are now a little bit different from the case of ideal
measurements as discussed in Section 2.
Since $1 - q$ is close, but not equal, to $p$ there is now
a small asymmetry between light and dark periods. 

In spite of the problems arising for $\Delta t\to0$ this limit can be
performed. The result is (details can again be found in \cite{BeHe3})
\begin{displaymath}
\lim_{\Delta t \to 0} \overline T_{\rm D} =
{\Omega_3^2 \over \Omega_2^2 A_3 } \;,\quad
\lim_{\Delta t \to 0} \overline T_{\rm L} = 
{\Omega_3^2(A_3^2+2\Omega_3^2) \over \Omega_2^2 A_3^3 }
\end{displaymath}
up to terms of relative order
$\mbox{\boldmath $\epsilon$}/\Omega_2 \tau_{\rm p}$.
In contrast to Eq.~(\ref{mean}) for ideal measurements
$\overline T_{\rm D}$ and $\overline T_{\rm L}$ remain finite, 
as physically expected.
For $\Delta t=0$ both driving fields are continuously on.
In this case the existence of light and dark periods is well
known under the name  ``electron shelving'' \cite{Dehmelt},
for which the same results for
$\overline T_{\rm D}$ and $\overline T_{\rm L}$ have been obtained
\cite{CT}.
\vspace*{0.5cm}

\noindent {\bf 6. Conclusions}

\vspace*{0.5cm}

The predictions of the quantum Zeno effect for a single system 
and an ensemble under rapidly repeated measurements have been studied
for the  example of state measurements on a two-level atom.
To test the quantum Zeno effect an experiment on an ensemble of atoms was 
performed by Itano et al.\ in which an atomic level measurement was
realized by means of a short laser pulse. 

An explanation for the approximately allowed
applicability of the projection postulate to this case has been 
given by us  using the quantum jump approach. We have determined
corrections to the ideal case explicitly.  
We have used these results  
to discuss the experiment of Itano et al.~\cite{Wine}
and a new possible experiment with a single atom in some detail. The 
projection postulate has been found to be an excellent pragmatic tool
for quick and fairly accurate answers and for a simple intuitive
understanding. However, corrections to it arise.

In the Introduction we have mentioned a controversy regarding the
quantum Zeno effect in general and the role of the experiment of Ref.
\cite{Wine} in particular. We think that our analysis sheds some
light on this. It is, in our opinion, perfectly legitimate to take
a `puristic' view that for  example the laser  pulses (``measuring pulses") 
have nothing to
do with  measurements but just lead to additional terms in the
Hamiltonian. Then any change in the temporal development is not
surprising and may in principle be calculated with these additional
interaction terms. However, the actual temporal behavior is in
general not easily seen and will often need numerical evaluations
which may give little physical insight. 
The other, more fruitful, point of view is that these 
laser pulses {\em approximately} realize measurements with
state reductions. Then one immediately has simple predictions for
the approximate behavior of the system and understands the
slow-down of the time evolution without complicated 
calculation. Finer details require of
course a finer analysis. An
actual freezing of the state does not seem possible since all
realistic measurements take a finite time. In the present case this
is explicitly seen in the finite duration of the laser pulse and the
required minimal time between them.

In a broader sense our analysis  also sheds some light on the use of the
projection postulate in general, not only in
connection with the Zeno effect. It seems that quite often the
projection postulate is a useful tool  which can give quick and
fairly accurate answers. The accuracy depends on how far the
particular realistic measurement differs from an ideal measurement as
considered in orthodox quantum mechanics, and corrections may have to
be taken into account. An idealization of realistic measurements and
the projection postulate may often be very useful. Over-idealizations,
however, are dangerous  since they can lead to interpretational
difficulties and to `paradoxes' like the
freezing of states in the limit of `continuous ideal measurements'.
In this limit the idealization becomes an overidealization and breaks down.

{\bf Acknowledgments.} This paper is based on talks given in 
Tayuan (Shanxi, China, June 1995) and in Goslar (Germany, July 1996). 
In China one of us (G.C.H.)
had the opportunity to enjoy stimulating discussions with Professor
Namiki on the quantum Zeno effect.

\newpage
\begin{figure}
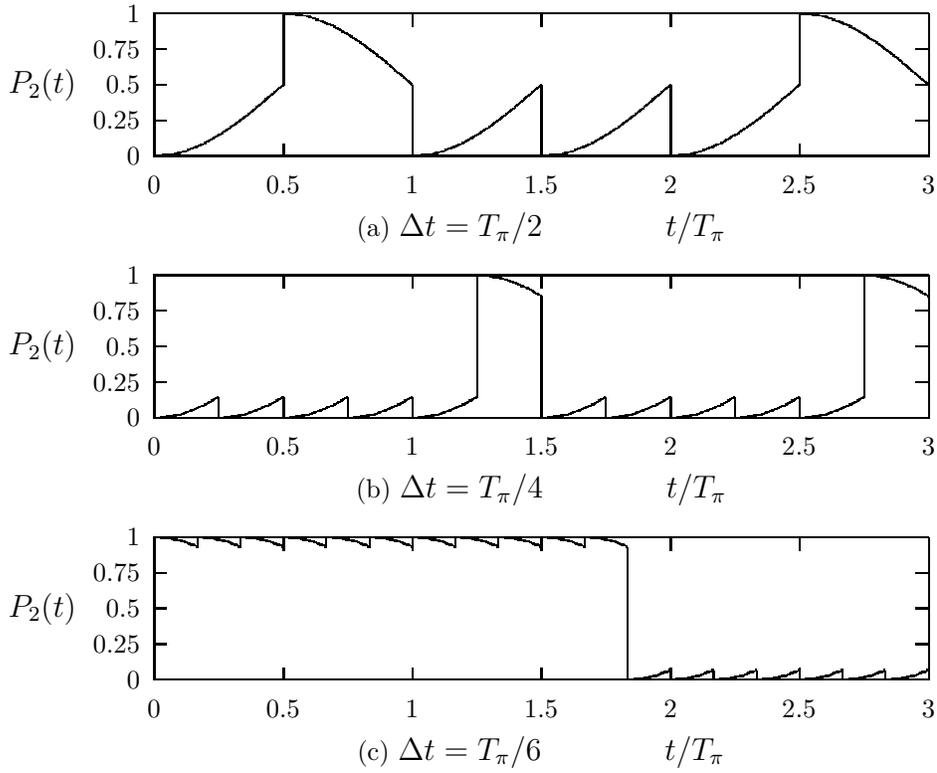

\setlength{\unitlength}{0.240900pt}
\ifx\plotpoint\undefined\newsavebox{\plotpoint}\fi
\sbox{\plotpoint}{\rule[-0.200pt]{0.400pt}{0.400pt}}%
\hspace*{0.6cm} 

\vspace*{0.4cm}
{\caption{Possible paths of the population $P_2(t)$  of level 2
of a single atom. Ideal measurements at times $\Delta t$ apart project
the atom on state $|1\rangle$ or $|2\rangle$ respectively. The time is
given in multiples of the length of a $\pi$ pulse $T_\pi$ with 
$T_\pi =\pi/\Omega_2$.}}
\end{figure}

\begin{figure}
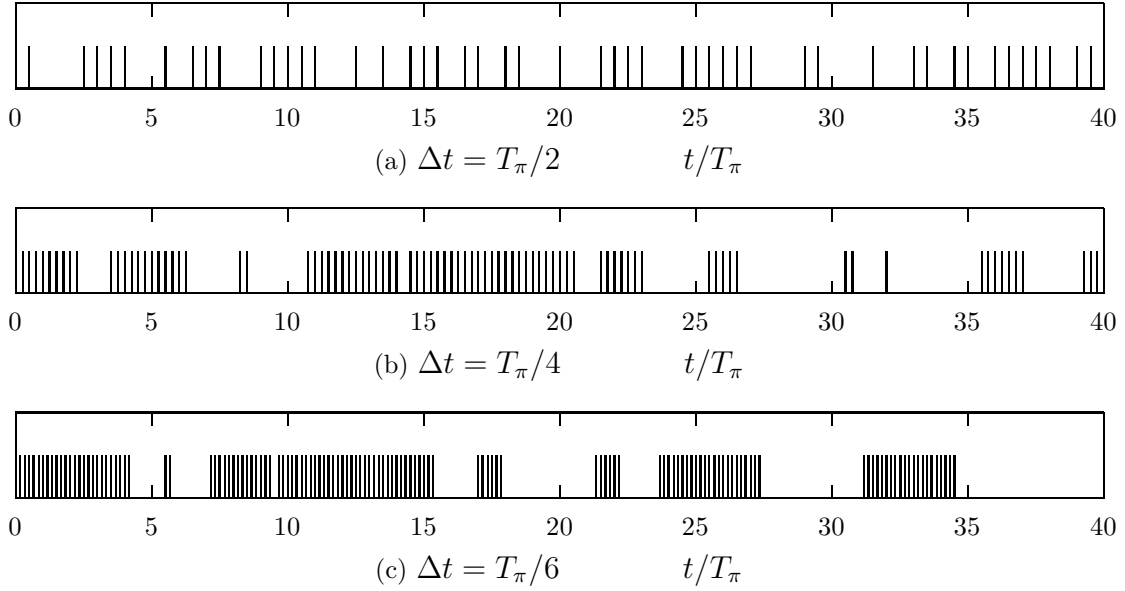

\setlength{\unitlength}{0.240900pt}
\ifx\plotpoint\undefined\newsavebox{\plotpoint}\fi
\sbox{\plotpoint}{\rule[-0.200pt]{0.400pt}{0.400pt}}%
\hspace*{0.8cm} %
{\caption{Stochastic alternating light and dark periods. The lines
mark times when the atom is found in state $|1\rangle$, with
accompanying light signal. 
$T_\pi=\pi/\Omega_2$ is the duration of a $\pi$ pulse.}}
\end{figure}
 
\begin{figure}
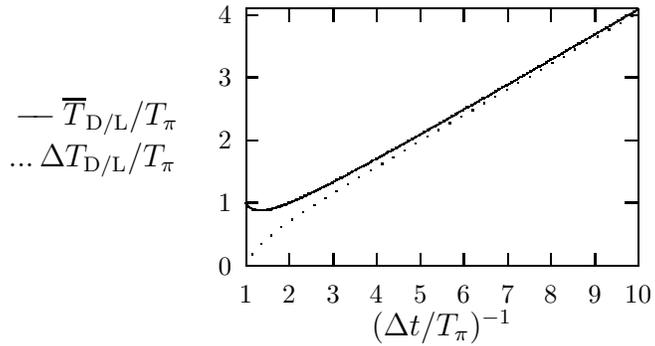

\setlength{\unitlength}{0.240900pt}
\ifx\plotpoint\undefined\newsavebox{\plotpoint}\fi
\sbox{\plotpoint}{\rule[-0.200pt]{0.400pt}{0.400pt}}%
\hspace*{1.9cm} %
{\caption{Mean length $\overline{T}_{\rm D/L}$ and standard deviation  
(dotted line) of light (dark) periods as a function of the inverse 
of the time between two measurements $\Delta t$. $T_\pi=\pi/\Omega_2$ 
is the duration of a $\pi$ pulse.}}
\end{figure}

\begin{figure}
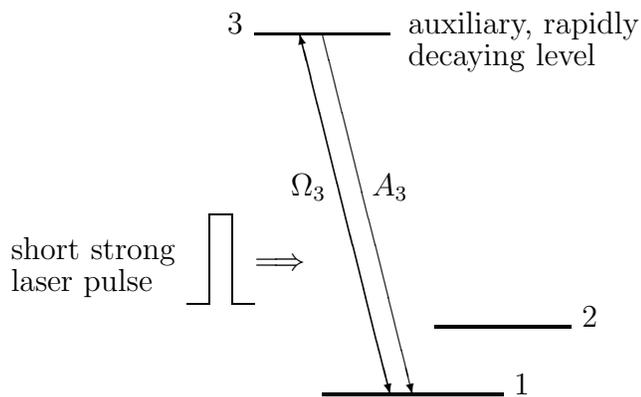

\hspace*{1cm}
\unitlength 0.6cm
%
\caption{V system with (meta)stable level 2 and Einstein coefficient
$A_3 $ for level 3. $\Omega_3$ is the Rabi frequency of the short
strong laser pulse of duration $\tau_{\rm p}$.
The 1--2 transition is not driven here.}
\end{figure}

\begin{figure}
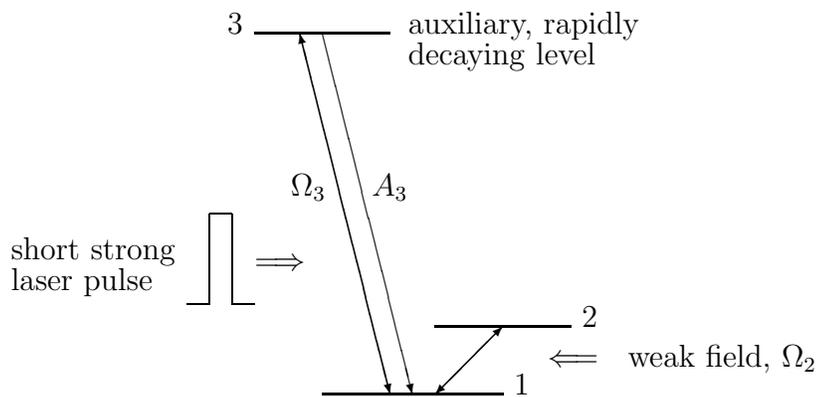

\hspace*{1cm}
\unitlength 0.6cm
%
\caption{V system as in Fig.~4 with an additional weak field
(Rabi frequency $\Omega_2$).}
\end{figure}

\begin{figure}
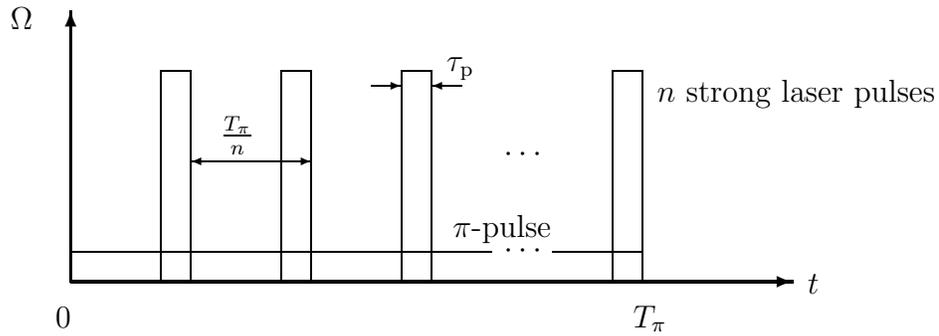

\unitlength 0.8cm
%
{\caption{Strong laser pulses and $\pi$ pulse as applied in the
 experiment of Itano et al. Initially all atoms are prepared in the
 ground state. $T_\pi$ is the duration of a $\pi$ pulse.}}
\end{figure}

\begin{figure}
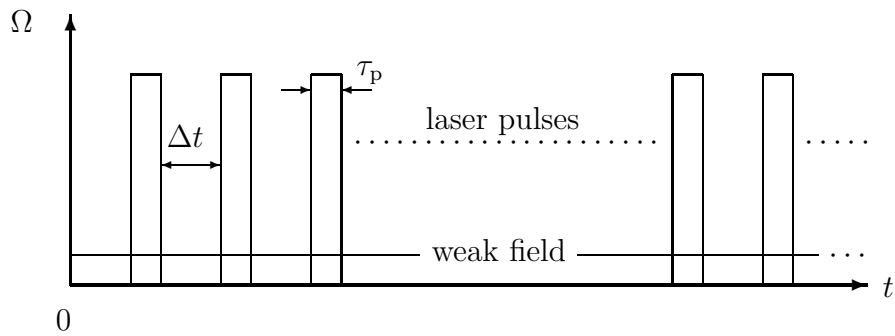

\unitlength 0.8cm
%
{\caption{Proposed experiment on a single atom. The weak field 
driving the 1--2 transition is kept on continuously. 
At times $\Delta t$ apart strong laser pulses are applied.}}
\end{figure}

\end{document}